\documentclass[12pt]{iopart}

\usepackage{graphicx}

\begin{document}

\title{Dynamic relaxation of topological defect at Kosterlitz-Thouless phase transition}

\author{X. P. Qin$^{1,2}$, B. Zheng$^{1}$ and N. J. Zhou$^{1}$}

\address{
$^1$ Department of Physics, Zhejiang University, Hangzhou 310027,
P.R. China\\
$^2$ School of Science, Zhejiang University of Science and Technology, Hangzhou 310023, P.R. China}
 \ead{\mailto{zheng@zimp.zju.edu.cn}}

\begin{abstract}
With Monte Carlo methods we study the dynamic relaxation of
a vortex state at the Kosterlitz-Thouless phase transition of the two-dimensional
\emph{XY} model. A local pseudo-magnetization is introduced to characterize the symmetric structure
of the dynamic systems.
The dynamic scaling behavior of the pseudo-magnetization and Binder cumulant
is carefully analyzed, and the critical exponents are determined.
To illustrate the dynamic effect of the topological defect, similar analysis for the
the dynamic relaxation with a spin-wave initial state is also performed for comparison.
We demonstrate that a limited amount of quenched disorder in the core of the vortex state
may alter the dynamic universality class.
Further, theoretical calculations based on the long-wave approximation are presented.
\end{abstract}


\maketitle

\section{Introduction}

In recent years many activities have been devoted to the study of
dynamic processes far from equilibrium. Compared with
spin-glass and structural-glass dynamics, critical dynamics
at standard second-order or Kosterlitz-Thouless (KT) phase transitions
is relatively simple. One is able to systematically explore the universal
dynamic scaling behavior far from equilibrium, up to
the {\it macroscopic} short-time regime
\cite{jan89,hus89,zhe98,zhe99}.
Although the spatial correlation length is still short in the beginning of the time
evolution, the dynamic scaling form is induced by the divergent
correlating time. Based on the
short-time dynamic scaling form, new methods for the determination
of both dynamic and static critical exponents as well as the
critical temperature have been developed
\cite{zhe98,zhe99,luo98}. Since the measurements are carried
out in the short-time regime, one does not suffer from critical
slowing down. Such a dynamic approach is especially useful, when one is interested
in tackling {\it both the statics and dynamics} of the critical systems.
Recent progress in the short-time critical dynamics
includes, for example, theoretical calculations and numerical
simulations of the {\it XY} models and Josephson junction arrays
\cite{zhe03a,oze03b,gra05,nie06,bri10}, magnets with quenched disorder
\cite{yin05,oze05,liu08,yot09,pru10}, ageing phenomena
\cite{god02,hen04,cal05,lei07,wal09}, domain-wall dynamics
\cite{zho08,he09,zho09,zho10}, weak first-order phase
transitions \cite{sch00,mon01,yin05,los09}, and various
applications and developments
\cite{alb02,lee05,ara06,lin08,sar09,hua10,puz10,zho10a}.

In the understanding of the critical dynamics far from equilibrium,
the dependence of the dynamic scaling behavior on the {\it
macroscopic} initial condition \cite{jan89,zhe98,fed06} is essential.
Up to now,
the dynamic relaxation with disordered and ordered initial states
has been systematically investigated \cite{jan89,luo98,zhe98,fed06}.
Physically, the disordered and ordered initial states correspond to the states at very high and zero temperatures respectively.
The ordered initial state can be also considered as a state under a strong external magnetic field.
Recently, the dynamic relaxation of a domain wall
is also concerned \cite{zho08,he09}, and it
shares certain common features with those around free and disordered
surfaces \cite{zho08}. The three relaxation processes above are
typical but relatively simple. By these successes, we are encouraged to tackle more general
relaxation processes.
On the other hand, for strongly disordered systems such as spin glasses, not only the metastable states but also the ground states are not simply
ordered and homogeneous, and the standard magnetization does not characterize the relaxation dynamics starting from a zero or low temperature.
Methodologically and technically one needs to develop new concepts.

From theoretical view points, metastable states and low-energy
excited states may play important roles in phase transitions.
One example is the spin-glass transition, and another
is the KT transition in the two-dimensional (2D) {\it XY} model. In the latter case,
low-energy excited states such as spin waves destroy the magnetic order,
and metastable state such as vortices and vortex pairs dominate
the phase transition.
Since vortices and vortex pairs are {\it topological defects},
the KT transition is sometimes called a topological phase transition.
The effect of topological defects is often a focus issue in the study of phase transitions, also in non-equilibrium dynamics \cite{bra00}.
In fact, one has not achieved too much understanding for
the dynamic relaxation of metastable states and low-energy excited states,
although its importance is often qualitatively addressed.

Recently many activities have been devoted to the study of vortex states and vortex structures
in nanowires and nanomagnets.
Attention is also drawn to the single vortex state
in experiments and numerical simulations \cite{che06,lua08,nag08,nog09,mas10}.
For example, a single vortex structure is explicitly observed
in the domain wall of a ferromagnetic nanowire \cite{nag08},
and a second-order transition between a domain wall and a vortex state
is detected \cite{mas10}. In Ref. \cite{nog09}, the abnormal diffusion
of a single vortex in the 2D {\it XY} model is also numerically simulated.
Therefore, the vortex state is a physically important object.
In this paper, we numerically and theoretically study the relaxation dynamics of a single vortex state
at the KT phase transition of the 2D {\it XY} model, in comparison with that of
a spin-wave state. The dynamic effect of the topological defect will be emphasized.
We demonstrate that it is essential to introduce
a local pseudo-magnetization to characterize the symmetric structure
of the dynamic systems, which
is defined as the projection of the spin configuration onto the metastable state or low-energy excited state.
In Sec. 2 the model and scaling analysis are described, and in
Sec. 3 the numerical results are presented. In Sec. 4 theoretical calculations based on the linearized long-wave approximation are presented.
Sec. 5 includes the conclusions.

\section{Model and scaling analysis}

\subsection{Model}
The 2D \emph{XY} model is the simplest model for
magnetic materials, exhibiting a KT phase
transition. The Hamiltonian is written as
\begin{equation}
-\frac{1}{kT}H=K\sum_{<ij>}\vec{S}_{i}\cdot\vec{S}_{j},
\label{equ1}
\end{equation}
where $\vec{S}_{i}=(S_{i,x},S_{i,y})$ is a planar unit
vector at site $i$ of a square lattice, the sum is over the nearest
neighbors, and $T$ is the temperature. In our notation, we simply take $K=1/T$. In the literatures, the
transition temperature $T_{KT}$ is reported to be between 0.89 and
0.90 for the 2D \emph{XY} model \cite{gup92,zhe99b,tom02}. Below
$T_{KT}$, the system remains critical. To study the pure relaxation dynamics of model A,
we adopt the "heat-bath"
algorithm of an one-spin flip, in which a trial move is accepted
with probability 1/[1+exp$(\Delta E/T)]$, where $\Delta E$ is the
energy change associated with the move.

\begin{figure}[ht]
\setlength{\unitlength}{1.2cm}
\begin{picture}(6,6)(0,0)
\includegraphics[width=7.cm]{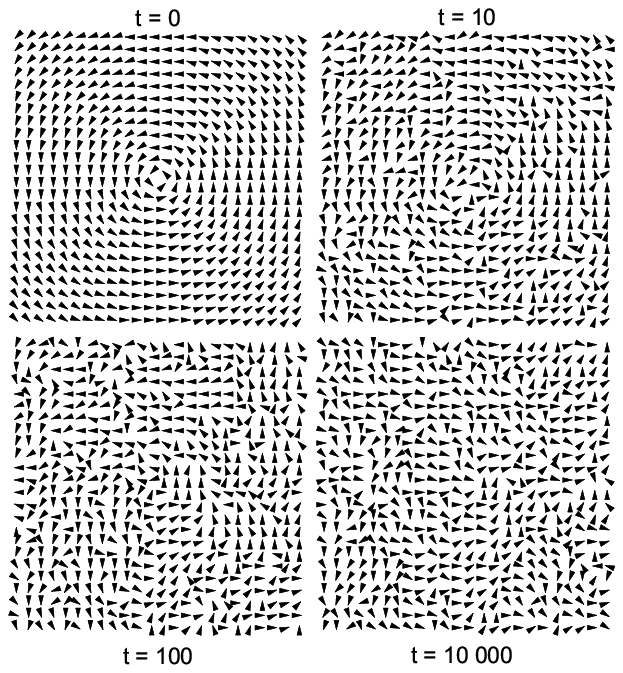}\quad\qquad
\includegraphics[width=7.2cm]{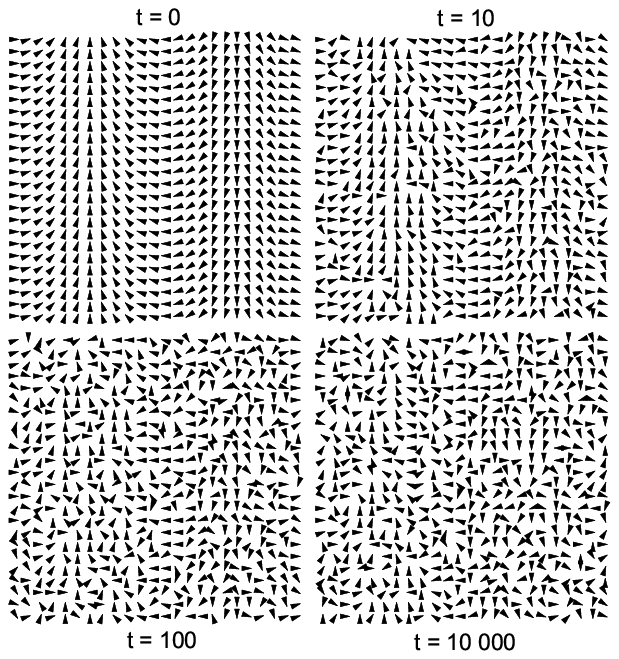}
\put(-12.5,-0.){\footnotesize(a)}
\put(-6.3,0.1){\footnotesize(b)}
\end{picture}
\caption{Dynamic relaxation of a vortex state and a spin-wave state is displayed respectively in (a) and (b),
for the $2$D {\it XY} model at the temperature T=0.89, slightly below $T_{KT}$.
The arrowhead denotes the orientation of the spin.
} \label{evolution}
\end{figure}

To study the dynamic relaxation of a single vortex state, we first
construct a {\it perfect} vortex as the initial state.
We put the vortex on a square lattice $L\times L$, and choose the center of the vortex
as the origin of the polar coordinates. To reduce the finite-size effect, we apply antisymmetric
boundary conditions with respect to the origin,
i.e., $\vec{S}_{\vec r}=-\vec{S}_{-\vec r}$, which mimic the vortex line on the boundary. In
Fig.~\ref{evolution}(a), the dynamic evolution of the spin
configuration is illustrated. Due to the vortex initial state, the
dynamic system is {\it inhomogeneous} for different radius $r$.
After the statistical average, the orientations of the spins form circular vortex lines.
Except for the topological defect at the center, the vortex
state is locally ordered, although globally it is still different
from an ordered state. Naively, we may measure the magnetization
along the circular vortex lines as a function of $r$ and $t$. From the
symmetry of the initial state, however, both the $x$ and $y$ components of
the magnetization are zero.

In order to characterize the dynamic relaxation of the vortex
state, therefore, we should introduce a {\it local
pseudo-magnetization} and its second moment by the projection to the
metastable state,
\begin{equation}
M_p^{(k)}(t,r)=\frac{1}{N_r^{k}}\left\langle\left
[\sum^{N_r}_{\varphi =1}\vec{S}_{r\varphi}(t)
\cdot\vec{S}_{r\varphi,p}\right ]^{k}\right\rangle, \quad
k=1,2; \label{equ4}
\end{equation}
\begin{equation}
M_n^{(k)}(t,r)=\frac{1}{N_r^{k}}\left\langle\left
[\sum^{N_r}_{\varphi =1}\vec{S}_{r\varphi}(t)
\cdot\vec{S}_{r\varphi,n}\right ]^{k}\right\rangle, \quad
k=1,2, \label{equ5}
\end{equation}
where $\langle \cdots \rangle$ represents the statistical average, and $N_r \approx 2\pi r$ is the number of the lattice sites at radius $r$.
In the case now, the metastable state is just the perfect vortex, the same as the initial state, $\vec{S}_{r\varphi,p}=\vec{S}_{r\varphi}(0)=\vec e_\varphi$,
while $\vec{S}_{r\varphi,n}=\vec e_r$ denotes the unit vector {\it
perpendicular} to $\vec{S}_{r\varphi,p}$, i.e., that of
the normal direction. In other words, $M_p^{(k)}(t,r)$ is the
$k$-th moment of the tangent direction, and $M_n^{(k)}(t,r)$ is that
of the normal direction. From the symmetry of the vortex initial
state, the normal component of the pseudo-magnetization is zero, i.e.
$M_n(t,r)=0$. Denoting $M(t,r)\equiv M_p (t,r)$ and
$M^{(2)}(t,r)\equiv M_p^{(2)}(t,r)+M_n^{(2)}(t,r)$, we then
define a time-dependent Binder cumulant,
\begin{equation}
U(t,r) = \frac {M^{(2)}(t,r)}{M(t,r)^{2}} - 1. \label{equ6}
\end{equation}
The Binder cumulant $U(t,r)$ describes the fluctuation of the
pseudo-magnetization.

To investigate the dynamic effect of quenched disorder, we may randomly fix the orientations
of the four spins at the center of the vortex
during the dynamic evolution. Such a dynamic system is topologically similar to
the dynamic relaxation around a disordered surface \cite{zho08,lin08}.
Experimentally, a vortex state may be prepared by applying an electric current
perpendicular to the lattice plane.

For comparison, we now consider the dynamic relaxation starting from
a low-energy excited state, i.e., a spin-wave state.
The spin wave is linearly polarized in the $x$ direction, while ordered in the $y$ direction.
The simulation is performed with a rectangle lattice $L_x\times L_y$. In
Fig.~\ref{evolution}(b), the dynamic evolution of the spin
configuration is displayed. The
spin-wave state is locally ordered, but globally not. For such a
dynamic process, we may naively measure the {\it line} magnetization along the $x$ direction
as a function of $t$. Obviously,
the magnetization is zero due to the periodicity of the spin-wave state.

In order to characterize the dynamic relaxation starting from
the spin-wave state, we may also define a pseudo-magnetization and
its second moment as functions of $t$,
\begin{equation}
M_p^{(k)}(t)=\frac{1}{L_x^{k}}\left\langle\left [\sum^{L_x}_{x
=1}\vec{S}_{xy}(t) \cdot\vec{S}_{xy,p}\right
]^{k}\right\rangle, \quad k=1,2; \label{equ9}
\end{equation}
\begin{equation}
M_n^{(k)}(t)=\frac{1}{L_x^{k}}\left\langle\left [\sum^{L_x}_{x
=1}\vec{S}_{xy}(t) \cdot\vec{S}_{xy,n}\right
]^{k}\right\rangle, \quad k=1,2, \label{equ10}
\end{equation}
where $\vec{S}_{xy,p}=\vec{S}_{xy}(0)$, $\vec{S}_{xy,n}$ denotes the unit vector
perpendicular to $\vec{S}_{xy,p}$, and $\langle \cdots \rangle$ represents
both the statistical average and the average in the $y$ direction. 
As shown in Fig.~\ref{evolution}(b), $\vec{S}_{xy}(0)$ periodically changes its orientation along the $x$ direction, but independent of $y$.
From the periodicity
of the spin-wave state, $M_n(t)=0$. Denoting $M(t)\equiv
M_p (t)$ and $M^{(2)}(t)\equiv
M_p^{(2)}(t)+M_n^{(2)}(t)$, we then define a time-dependent
Binder cumulant,
\begin{equation}
U(t) = \frac {M^{(2)}(t)}{M(t)^{2}} - 1. \label{equ11}
\end{equation}

\subsection{Scaling analysis}

In the critical regime, there are three spatial length scales in the
dynamic relaxation of the vortex state, i.e., the nonequilibrium
spatial correlation $\xi(t)$, the radius $r$ of the circular vortex line, and the
lattice size $L$. Therefore general scaling arguments lead to the
scaling forms of the magnetization and its second moment,
\begin{equation}
M^{(k)}(t,r,L)=\xi(t)^{-k\eta /2}
\widetilde{M}^{(k)}(r/\xi(t),\xi(t)/L),\quad k = 1,2, \label{equ12}
\end{equation}
where $\eta$ is the static exponent. In the
equation, the factor $\xi(t)^{-k\eta /2}$ indicates the
scaling dimension of $M^{(k)}$, and the scaling function
$\widetilde{M}^{(k)}(r/\xi(t),\xi(t)/L)$ represents the scale
invariance of the dynamic system. In general, we expect that the
scaling forms in Eq.~(\ref{equ12}) holds already in the
$macroscopic$ short-time regime, after a microscopic time scale
$t_{mic}$ \cite{jan89,zhe98}. Our numerical simulations show,
however, that only the pseudo-magnetization and its second moment
obey such scaling forms.

In the short-time regime, i.e., the regime with $\xi(t)\ll L$, the
pseudo-magnetization is independent of $L$. Then the scaling form is
simplified to
\begin{equation}
M(t,r)=\xi(t)^{-\eta /2} \widetilde{M}(r/\xi(t)). \label{equ13}
\end{equation}
Since the spatially correlating terms in the susceptibility $M^{(2)}(t,r)-M(t,r)^2$
can be neglected for $\xi(t)\ll r$, one may deduce $U(t,r)\sim 1/N_r^{d-1}$
for large $r$, with $d=2$. Here $N_r \propto r$ is the number of the lattice sites at radius $r$,
similar to the lattice size $L$ in Refs.~\cite{zhe98, zho08,he09}.
Further, $U(t,r)$ is independent of the scaling
variable $\xi(t)/L$ in the short-time regime. Together with Eqs.~(\ref{equ12})
and (\ref{equ13}), one may write down the scaling form
\begin{equation}
U(t,r)=[\xi(t)/r]^{d-1} \widetilde{U}(r/\xi(t)).
\label{equ14}
\end{equation}
The Binder cumulant is interesting, for the static exponent $\eta$
is \emph{not} involved.

For the critical dynamics of a continuous phase transition, $\xi(t)$
usually grows by a power law $\xi(t)\sim t^{1/z}$, and $z$ is the
so-called dynamic exponent. In shorter times, there may be
corrections to scaling, typically in a power-law form
\begin{equation}
\xi(t) \sim t^{1/z}(1+c/t^b). \label{equ15}
\end{equation}
For magnetic systems with a KT phase transition, e.g., the 2D \emph{XY} model,
the correction to scaling is weak in the dynamic relaxation
with an \emph{ordered} initial state, the dynamic exponent is
theoretically expected to be $z=2$ and numerical simulations indicate $b \approx 1$ \cite{zhe03a,lei07}. Due to the dynamic
effect of the vortex-pair creation and annihilation, however, the correction to
scaling becomes strong in the dynamic relaxation with a
{\it disordered} initial state, and both theoretical and numerical calculations lead to a logarithmic form
\cite{bra00,bra00a,zhe03a,abr04,lei07}
\begin{equation}
\xi(t) \sim [t/(ln\emph{t}+c)]^{1/z}. \label{equ16}
\end{equation}
Theoretically, Eq.~(\ref{equ16}) is equivalent to Eq.~(\ref{equ15})
in the limit $b\rightarrow 0$.
Numerically detecting a logarithmic correction to scaling is rather
notorious, for it is negligible only in the limit
$t\rightarrow\infty$ \cite{zhe03a,abr04,lei07}.
Recently, a logarithmic correction to scaling is detected inside the domain interface,
which is also attributed to the vortex-pair creation and annihilation \cite{he09}.

In this paper, we will show that there exists a \emph{core} of the vortex state,
and there emerges a strong logarithmic correction to scaling in the core, again due to the vortex-pair creation and annihilation.
The core of the vortex grows with time, and also roughens. Such a phenomenon
is rather similar to the propagation and roughening of the domain
interface \cite{zho08,he09,zho09}. Outside the core of the vortex and also for the spin-wave initial state,
the dynamic system is locally ordered. Therefore,
the dynamic relaxation is similar to that with the ordered initial state, and the correction to scaling is in a weak power-law form.
Further, we demonstrate that a limited amount of quenched disorder
in the core of the vortex may change
the dynamic universality class, i.e., the scaling functions
$\widetilde{M}^{(k)}(s)$ in Eq.~(\ref{equ12}) and the corresponding critical exponents.

For the dynamic relaxation with the spin-wave initial state, there
are two spatial length scales in the dynamic system, i.e., the
nonequilibrium spatial correlation $\xi(t)$, and the lattice size
$L_x$, if the lattice size $L_y$ is sufficiently large.
Therefore scaling arguments lead to the scaling form of the
pseudo-magnetization and its second moment,
\begin{equation}
M^{(k)}(t,L_x)=\xi(t)^{-k\eta /2}
\widetilde{M}^{(k)}(\xi(t)/L_x),\quad k = 1,2. \label{equ17}
\end{equation}
For the pseudo-magnetization, the scaling function
$\widetilde{M}(\xi(t)/L_x)$ is independent of $L_x$ in the
thermodynamic limit $L_x\rightarrow \infty$. The scaling form is
simplified to
\begin{equation}
M(t) \sim \xi(t)^{-\eta /2}. \label{equ18}
\end{equation}
The scaling behavior of the Binder cumulant is,
\begin{equation}
U(t)\sim [\xi(t)/L_x]^{d-1}. \label{equ19}
\end{equation}

\section{Monte Carlo Simulation}

For the dynamic relaxation with the vortex initial state, our main
results are obtained with $L=256$ at $T_{KT}=0.89$, and the maximum
updating time is $t_M=10 240$ Monte Carlo time steps. A Monte Carlo time step is defined
by a sweep over all spins on the lattice.
Additional simulations with $L=128$ and $L=512$ are performed, to investigate
the finite-size scaling behavior and finite-size effect. Total
samples for average are about $20 000$. For the dynamic relaxation
with the spin-wave initial state, the main results are obtained with
$L_x\times L_y=256\times 256$ and $L_x\times L_y=512\times 256$ at
$T_{KT}=0.89$, and the maximum updating time is $t_M=10 240$ Monte
Carlo time steps. Additional simulations with $L_x\times L_y=128\times
256$ and $L_x\times L_y=1024\times 256$ are also performed. Total
samples for average are about $10 000$. Statistical errors are
estimated by dividing the total samples into two or three subgroups.
If the fluctuation in the time direction is comparable with or
larger than the statistical error, it will be taken into account.
Theoretically, the scaling forms described in the previous section
hold in the macroscopic short-time regime, after a microscopic time
scale $t_{mic}$. $t_{mic}$ is not universal, and relies on
microscopic details of the dynamic systems. In our simulations,
$t_{mic}$ is about $100$ Monte Carlo time steps.

\subsection{Vortex state}

\begin{figure}[p]
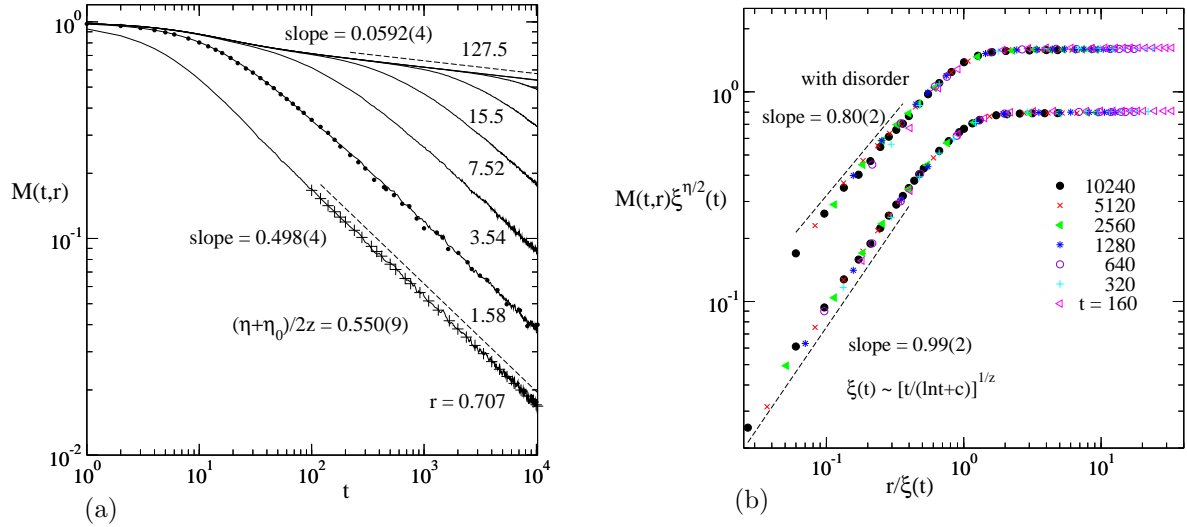

\setlength{\unitlength}{1.2cm}
\begin{picture}(6,6)(0,0)
\includegraphics[width=7.2cm]{Mtr.eps}\quad\quad
\includegraphics[width=7.6cm]{Mr_dc.eps}
\put(-12.2,-0.2){\footnotesize(a)}
\put(-5.,-0.1){\footnotesize(b)}
\end{picture}
\caption{(a) The time
evolution of the pseudo-magnetization of the $2$D {\it XY} model starting from
the vortex state is displayed for different radius $r$ on a double-log scale.
Simulations are performed with $L=256$, and solid dots are the data obtained with $L=512$.
Dashed lines represent power-law fits,
and pluses indicate the fit with a logarithmic correction to scaling. (b) The
scaling function $\widetilde{M}(r/\xi(t))=M(t,r)\xi(t)^{\eta/2} $ is plotted on a double-log scale.
In the core of the vortex, the logarithmic correction to scaling is taken into account, and $c=5.45$
is from the fitting in (a). Data
collapse is observed for different $t$. The dashed line shows a power-law fit.}
\label{mtr}
\end{figure}

For the dynamic relaxation with the vortex initial state, the time evolution of the local
pseudo-magnetization $M(t,r)\equiv M_p^{(1)} (t,r)$ defined in
Eq.~(\ref{equ4}) is displayed in Fig.~\ref{mtr}(a).
Monte Carlo simulations have been performed with the lattice sizes
$L=256$ and $512$, and the finite-size effect is negligibly small. Denoting
$s=r/\xi(t)$, we observe that for a sufficiently large $s$, e.g.,
$r=127.5$ and $t < t_M=10 240$, $M(t,r)$ approaches the power-law
decay \emph{at bulk}, i.e., $M(t,r) \sim t^{-\eta/2z}$, while for a sufficiently small $s$, e.g.,
$r=0.707$ and $t> 100$, $M(t,r)$ also exhibits a power-law behavior,
but decays \emph{much faster} than at bulk. In other words, the
scaling function $\widetilde{M}(s)$ in Eq.~(\ref{equ13}) is
characterized by
\begin{equation}
   \widetilde{M}(s) \sim \{
   \begin{array}{lll}
    \mbox{const}    & \quad &  \mbox{$s \to  \infty$ } \\
    s^{\eta_0 /2}  & \quad &  \mbox{$s \to  0$}
   \end{array}.
\label{equ20}
\end{equation}
Here $\eta_0$ is the vortex exponent. We call the regime with small $s$ the {\it core} of the vortex state.
The spatial length scale of the core increases with time,
and it is just proportional to the nonequilibrium correlation length $\xi(t)$.
Outside the core of the vortex, i.e. the regime with large $s$,
the dynamic relaxation of the pseudo-magnetization is governed by the bulk exponent $\eta$, while
in the core of the vortex, it is controlled by
both the bulk exponent $\eta$ and the vortex exponent $\eta_0$.
Outside the core of the vortex, the dynamic relaxation of the pseudo-magnetization is
the same as that with an ordered initial state, and the correction
to scaling is weak. Assuming $\xi(t)\sim t^{1/z}$, one deduces
$M(t,r)\sim t^{-\eta/2z}$ for $s\rightarrow \infty$. In Fig.~\ref{mtr}(a),
the exponent $\eta/2z=0.0592(4)$ measured from the slope of the
curve of $r=127.5$ is in agreement with that for the ordered initial
state \cite{zhe03a,he09}, and it is also consistent with $\eta=0.234(2)$ and $z=2$
reported in the literatures \cite{zhe03a}. A power-law correction to
scaling slightly refines the result by about one percent, and yields $\eta/2z=0.0586$.
In fact, the exponent $\eta$ extracted in almost all the numerical simulations is somewhat 
smaller than the prediction of the KT theory, $\eta=1/2$. This may imply that the KT theory is not exact for the lattice {\it XY} model. 
In the core of the vortex, one may similarly derive $M(t,r)\sim t^{-(\eta+\eta_0)/2z}$ for
$s\rightarrow 0$. Then one measures $(\eta+\eta_0)/2z=0.498(4)$ from
the slope of the curve of $r=0.707$, and calculates
$\eta_0/2=0.878(8)$ by taking $z=2$ as input.

For the domain interface of the 2D \emph{XY} model, the interface
exponent $\eta_0/2$ is reported to be 0.997(7), very close to 1
\cite{he09}. It indicates that $M(t,r)$ is an analytic function of
$r$. Topologically, a vortex state could be transformed from either a domain interface or a free surface.
Physically, the former seems closer to a vortex state.
Is the vortex exponent $\eta_0/2$ really different from 1?
According to the scaling form in Eq.~(\ref{equ20}), the $r$-dependence and $t$-dependence of the magnetization $M(t,r)$
should yield a same exponent $\eta_0/2$.
We observe that at a fixed time $t$, the $r$-dependence of $M(t,r)$ is very close to linear in the small $r$ regime.
According to Eq.~(\ref{equ20}), this indicates $\eta_0/2 \approx 1$, contracting to $\eta_0/2=0.878(8)$ from the power-law fit
in Fig.~\ref{mtr}(a). Our thought is that there exists a strong
logarithmic correction in the growth law of $\xi(t)$,
described by Eq.~(\ref{equ16}). Such a fitting to the curve of
$r=0.707$ in Fig.~\ref{mtr}(a) yields $(\eta+\eta_0)/2z=0.550(9)$.
With $\eta/2z=0.0586$ as input, one then calculates the vortex exponent
$\eta_0/2=0.983(18)$, also close to 1. To further confirm this result,
let us fully examine the scaling form in Eqs.~(\ref{equ13}) and (\ref{equ20}).
In Fig.~\ref{mtr}(b), the scaling function
$\widetilde{M}(r/\xi(t))=M(t,r)\xi(t)^{\eta/2} $ is plotted,
and data collapse is clearly observed for different time $t$.
In the small-$s$ regime, the logarithmic correction to scaling described by Eq.~(\ref{equ16})
has been taken into account, and the constant $c=5.45$ is taken from
the fitting in Fig.~\ref{mtr}(a). Obviously, $\widetilde{M}(s)$
exhibits a power-law behavior in the small-$s$ regime, supporting the linear $r$-dependence of $M(t,r)$. From the slope of the curve of $t=10
240$, one measures ${\eta_0/2}=0.99(2)$, rather close to 1, and
consistent with $\eta_0/2=0.983(18)$ obtained from Fig.~\ref{mtr}(a).
For large $s$, $\widetilde{M}(s)$ approach a constant.
Finally, we mention that in the data collapse in Fig.~\ref{mtr}(b),
data points for all $r$ including $r=0.707$ are plotted, although those for $t \leq t_{mic} \sim 100$ or $200$ are truncated.
In other words, the dynamic scaling behavior holds even in the very center of the vortex state.

\begin{figure}[ht]
\setlength{\unitlength}{1.2cm}
\begin{picture}(6,6)(0,0)
\includegraphics[width=7.cm]{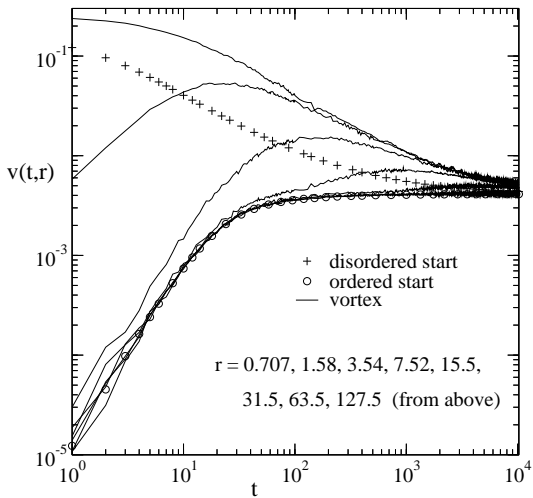}
\end{picture}
\caption{The time evolution of the vortex density in the dynamic
relaxation of a vortex state, in comparison with that in the
dynamic relaxation starting from ordered and disordered states.
} \label{vortex}
\end{figure}

The logarithmic correction to scaling at the KT phase transition is
believed to be induced by the vortex-pair creation and annihilation. Therefore we
measure the time evolution of the vortex density for different $r$,
in comparison with those in the dynamic relaxation starting from
ordered and disordered states. The vortex density is defined as
\begin{equation}
v(t,r)=\langle|v_p|\rangle,
v_p=\sum_{p(r,\varphi)}[\theta_i(t)-\theta_j(t)]/2\pi, \label{equ21}
\end{equation}
where $\theta_i$ and $\theta_j$ denote the orientational angles of
$\vec{S}_i$ and $\vec{S}_j$,
$(\theta_i-\theta_j)$ are valued within the interval $[-\pi,\pi]$,
the sum is over the four links $(i,j)$ of the clockwise plaquette at
site $(r,\varphi)$, and $\langle \cdots \rangle$ represents both the statistical
average and the average in the $\varphi$ direction. Numerical
results are shown in Fig.~\ref{vortex}. Outside the core of the vortex,
e.g., at $r=127.5$, the vortex density is initially zero, then
increases with time, and finally reaches the steady value in
equilibrium. This dynamic process is relatively fast, the same as
that in the dynamic relaxation with the ordered initial state. In
the core of the vortex, e.g., at $r=0.707$ or $1.58$, the vortex density
initially increases rapidly to a large value, which even exceeds
that of the disordered initial state, then decreases slowly and
relaxes to the equilibrium. This dynamic process is even slower than
that with the disordered initial state. Therefore it is not
surprising that a strong logarithmic correction to scaling emerges.

Here we should emphasize that the slow dynamic relaxation in the core of the
vortex state is indeed induced by the topological structure.
For example, we may simulate the dynamic relaxation for the initial
state with a four-spin vortex on the ordered or disordered
background. The pseudo-magnetization rapidly drops to zero.

\begin{figure}[ht]
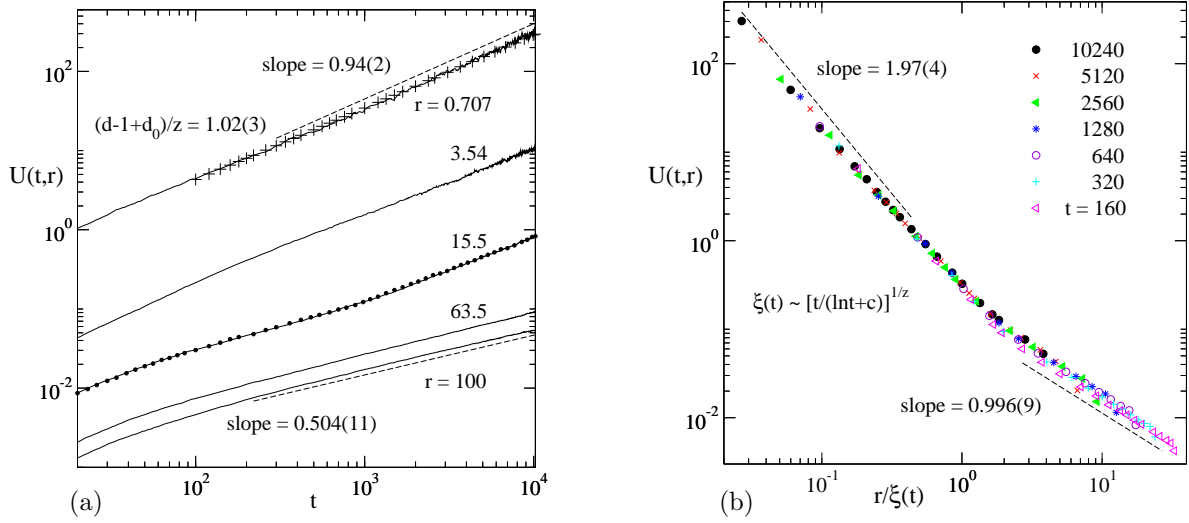

\setlength{\unitlength}{1.2cm}
\begin{picture}(6,6)(0,0)
\includegraphics[width=7.2cm]{Utr.eps}\quad\qquad
\includegraphics[width=7.2cm]{Ur_dc.eps}
\put(-12.4,-0.){\footnotesize(a)}
\put(-5.2,-0.){\footnotesize(b)}
\end{picture}
\caption{(a) The time evolution of the Binder cumulant of the $2$D {\it XY} model starting from the
vortex state is displayed for different radius $r$ on a double-log scale.
Simulations are performed with $L=256$, and solid dots are the data obtained with $L=512$.
Dashed lines represent power-law fits,
and pluses indicate the fit with a logarithmic correction to scaling. (b) Data
collapse of the Binder cumulant is displayed for different $t$. In the core of the vortex, the
logarithmic correction to scaling is taken into account, and $c=5.45$. Dashed lines show power-law fits.}
\label{utr}
\end{figure}

To clarify the scaling form of the Binder cumulant for the dynamic relaxation of a vortex state,
we plot $U(t,r)$ as a function of $t$ for
different radius $r$ in Fig.~\ref{utr}(a). Similar to
Eq.~(\ref{equ20}), we expect the scaling function $\widetilde{U}(s)$ in Eq.~(\ref{equ14}) to be
\begin{equation}
   \widetilde{U}(s) \sim \{
   \begin{array}{lll}
    \mbox{const}    & \quad &  \mbox{$s \to  \infty$ } \\
    s^{-d_0}  & \quad &  \mbox{$s \to  0$}
   \end{array}.
\label{equ22}
\end{equation}
In the core of the vortex state, e.g., $r=0.707$, $U(t,r)$
looks like exhibiting a power-law behavior, and the slope of the
curve is $0.94(2)$. Assuming $\xi(t)\sim t^{1/z}$, one derives
$U(t)\sim t^{(d-1+d_0)/z}$. From $(d-1+d_0)/z=0.94(2)$, one
calculates $d_0=0.88(4)$. With the logarithmic correction in
Eq.~(\ref{equ16}), however, the fitting yields $d_0=1.04(6)$, close to $1$.
To further confirm this result, we may fully examine the scaling form
in Eqs.~(\ref{equ14}) and (\ref{equ22}). In Fig.~\ref{utr}(b),
$U(t,r)$ is plotted as a function of $s=r/\xi(t)$. Data collapse
is observed for different time $t$. In the core of the vortex, the logarithmic correction to scaling
has been taken into account. Obviously, $U(t,r)$ exhibits a power-law behavior
in the core of the vortex. From the
slope of the curves, one measures $d-1+d_0=1.97(4)$, then calculates
${d_0}=0.97(4)$, and it supports that $d_0$ is close to 1.
For comparison, the exponent $d_0$ for the domain interface is close to 2 \cite{he09}.
The reason is that the domain interface is a one-dimensional object,
while the core of the vortex is effectively
zero-dimensional. It is interesting that the pseudo-magnetization of
the zero-dimensional vortex core exhibits a similar dynamic behavior as
that of the one-dimensional domain interface, but its fluctuation is
different. In fact, the fluctuation of the pseudo-magnetization at the core of the vortex state
does not show $r$-dependence. Therefore, we have the simple relation $d_0=\eta_0-(d-1)$.

Outside the core of the vortex, e.g., $r=100$, a
power-law behavior is observed in Fig.~\ref{utr}(a), and the slope of
the curve is $0.504(11)$. Assuming $\xi(t)\sim t^{1/z}$, one derives
$U(t,r)\sim t^{(d-1)/z}$. From $(d-1)/z=0.504(11)$, we obtain
$z=1.98(4)$, consistent with the theoretical value $z=2$. At the
same time, we may also examine the behavior $U(t,r)\sim
s^{-(d-1)}$ in the large-$s$ regime in Fig.~\ref{utr}(b).
From the slope of the curves, one measures
${d-1}=0.996(9)$. Since the maximum radius of the vortex line is $r=L/2$
in the square lattice, the dynamic behavior of the Binder cumulant
becomes unreasonable for $r \ge L/2$. Due to this boundary effect, the data collapse
of the Binder cumulant in Fig.~\ref{utr}(b) is less good for large $s$, compared with that for small $s$ and
for the magnetization shown in Fig.~\ref{mtr}(b).

\subsection{Vortex state with quenched disorder}

To study the dynamic effect of quenched disorder, and to further demonstrate the importance of the topological structure of the vortex state,
we randomly fix the orientations of the four spins at the center of the vortex,
and repeat the simulation of the dynamic relaxation of the vortex state and analysis of the pseudo-magnetization according to Eqs.~(\ref{equ13}) and (\ref{equ20}).
The scaling function $\widetilde{M}(r/\xi(t))$ is plotted in Fig.~\ref{mtr}(b),
and data collapse is also observed. However, the vortex exponent is modified to ${\eta_0/2}=0.80(2)$,
significantly different from ${\eta_0/2}=0.99(2)$ for the case without disorder. Such a scenario is similar to
the dynamic relaxation around a disordered surface \cite{zho08,lin08}. But here disorder exists only at four lattice sites in the core of the vortex state.

Topologically, the vortex state may be compared with a domain wall or a surface.
It is interesting that the dynamic relaxation of the vortex state is similar to that of
a domain wall rather than a surface, i.e., $\eta_0/2=1.0$.
However, a limited amount of quenched disorder
around the center of the vortex could modify the dynamic universality class,
leading to $\eta_0/2=0.80(2)$.
This again shows that the topological structure of the vortex state is crucial.

\subsection{Spin-wave state}

\begin{figure}[p]
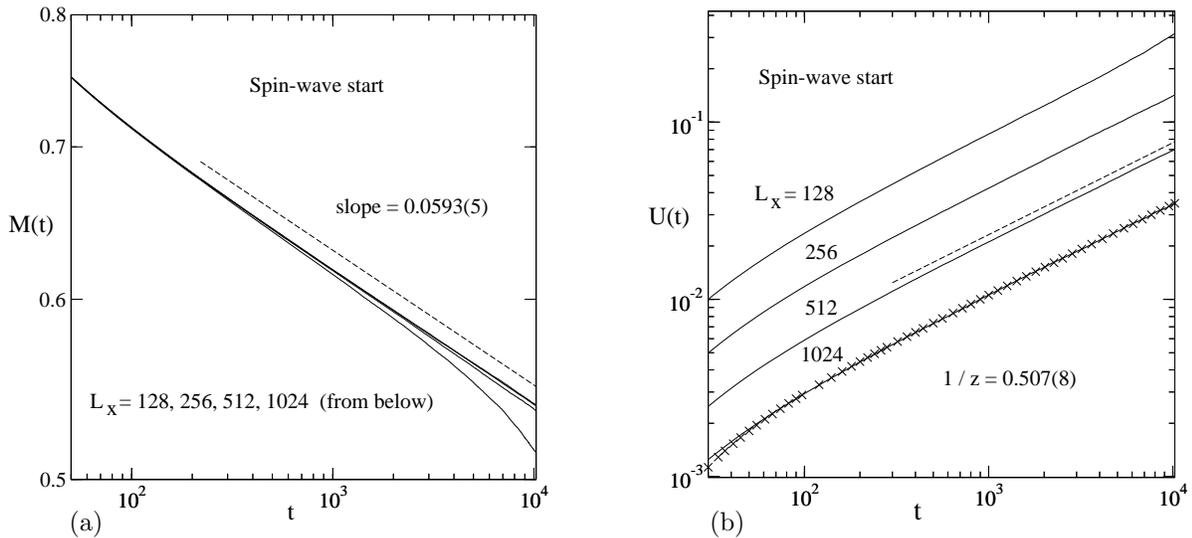

\setlength{\unitlength}{1.2cm}
\begin{picture}(6,6)(0,0)
\includegraphics[width=7.2cm]{MtSP.eps}\quad\qquad
\includegraphics[width=7.2cm]{UtSP.eps}
\put(-12.4,-0.1){\footnotesize(a)}
\put(-5.3,-0.1){\footnotesize(b)}
\end{picture}
\caption{In (a) and (b), the pseudo-magnetization and Binder cumulant of the $2$D {\it XY}
model starting from the spin-wave state are plotted on a double-log
scale. Dashed lines show power-law fits, while crosses indicate the fitting
with a power-law correction to scaling.} \label{mtsp}
\end{figure}

For the spin-wave initial state, the time evolution of the
pseudo-magnetization is displayed in Fig.~\ref{mtsp}(a). Since the
spin-wave state is locally ordered, the dynamic relaxation of the
pseudo-magnetization is expected to be similar to that with the ordered
initial state, $M(t)\sim t^{-\eta/2z}$, {\it without} an logarithmic correction to scaling.
In fact, if one computes the vortex density $v(t)$,
it completely overlaps with that of the dynamic relaxation starting from an ordered state as shown in Fig.~\ref{vortex}.
In Fig.~\ref{mtsp}(a), the
exponent $\eta/2z=0.0593(5)$ is estimated from the slope of the
curve of $L_x=512$ and $1024$. A power-law correction to scaling
refines the measurement to $\eta/2z=0.0586(3)$, with $b=1$ and $c=-11.1$.

For the dynamic relaxation with both the vortex and spin-wave
initial states, the standard magnetization is zero from the beginning
of the time evolution. Only the local pseudo-magnetization
could characterize the dynamic systems.

For the spin-wave initial state, the time evolution of the Binder
cumulant is displayed in Fig.~\ref{mtsp}(b). The dynamic relaxation of
the Binder cumulant is similar to that with an ordered initial
state. Assuming $\xi(t)\sim t^{1/z}$, one deduces $U(t)\sim
t^{(d-1)/z}$. In Fig.~\ref{mtsp}(b), the slope of the curves of
$L_x=512$ and $1024$ is estimated to be $0.520(3)$. Obviously, there exists a
correction to scaling in the growth law of $\xi(t)$,
and the fitting with Eq.~(\ref{equ15}) yields an exponent
$(d-1)/z=0.507(8)$, consistent with the theoretical value $z=2$.
Finally, the finite-size dependence $U(t) \sim 1/L_x^{d-1}$ can be also verified
with the data in Fig.~\ref{mtsp}(b).

\section{Theoretical calculation}

Based on the linearized long-wave approximation, we may theoretically derive the scaling forms
for the dynamic relaxation with the spin-wave and vortex initial state. 
Here we just report the main results, and leave the detailed calculations in the Appendix 1.
For the spin-wave initial state, the local pseudo-magnetization is independent of $\vec R$,
\begin{equation}
M_{p}(t, \vec R) \propto  t^{- \eta/2z}.
\label{equ2240}
\end{equation}
This power-law behavior is in agreement with the scaling form in Eq.~(\ref{equ18}).
For the vortex initial state, the scaling form of the pseudo-magnetization is
\begin{equation}
M_{p}(t,  {r'}) = t^{-\eta / 2z}\sin(\sqrt{\pi}
\int_{0}^{r'}dR'e^{-R'^2}), \label{equ2210}
\end{equation}
where $r' = r / \sqrt{4a^2t/T} \propto r/t^{1/z}=s$. This result is principally
in agreement with the scaling forms in Eqs.~(\ref{equ13}) and (\ref{equ20}), although the logarithmic correction is still absent
due to the linearized long-wave approximation in Eq.~(\ref{equ420}).
For small $r'$, $M_{p}(t, {r'}) \sim t^{-\eta / 2z} s$, therefore, $\eta_0/2=1$.

\section{Conclusion}

With Monte Carlo methods, we have simulated the dynamic relaxation
starting from both a vortex state and a spin-wave state
at the KT phase transition of the $2$D \emph{XY} model.
The dynamic scaling behavior of the local pseudo-magnetization and Binder cumulant
is carefully analyzed. Although the core of the vortex state is effectively zero-dimensional,
it exhibits a universal scaling behavior, controlled by both
the bulk exponent $\eta$ and the vortex exponent $\eta_0$.
Taking into account the logarithmic correction,
the vortex exponent $\eta_0/2=0.99(2)$ is extracted.
A limited amount of quenched disorder in the core of the vortex state alters the dynamic universality class,
and the vortex exponent is estimated to be $\eta_0/2=0.80(2)$.
In Table~\ref{t1}, all the measurements of the critical exponents are summarized, in comparison with those 
in the literatures.
Finally, theoretical calculations based on the long-wave approximation are presented,
and the scaling forms for the dynamic relaxation
with the spin-wave and vortex initial states are derived.

Our numerical and theoretical results show that
the local pseudo-magnetization is a useful concept in characterizing
the relaxation dynamics starting from a metastable state or a low-energy excited state.
Further applications of the methodology
to complex dynamic systems such as spin glasses are interesting.

\ack  This work was supported in part by NNSF of
China under Grant Nos. 10875102 and 11075137.

\section*{Appendix 1. Theoretical calculation}

In this appendix, we theoretically derive the scaling forms for the dynamic relaxation
with the spin-wave initial state and the vortex initial state, based on the linearized long-wave approximation.
In the Hamiltonian in Eq.~(\ref{equ1}), the spin $\vec S_i $ can be represented by
an angle, i.e., $\vec S_i = (\cos \theta_i, \sin \theta_i)$. With the linearized long-wave approximation, the Hamiltonian can be reduced,
\begin{eqnarray}
H &=& - \sum_{<ij>}\cos(\theta_i - \theta_j)
\approx  \frac{1}{2}\sum_{<ij>}\left(\theta_i - \theta_j \right)^2  + H_0\nonumber \\
& = &  \frac{1}{4}\sum_{\vec R, \vec a} \left(\theta(\vec R) -
\theta(\vec R + \vec a) \right)^2 +H_0,
\label{equ420}
\end{eqnarray}
where $\vec R$ is the position vector of a lattice site and
$\vec a$ is the spacing vector between the site and its nearest neighbors.
The constant $H_0$ will be ignored in the following calculations. After the
fourier transformation, the Hamiltonian is rewritten as
\begin{equation}
H = \frac{1}{2}\sum_{\vec k}J(\vec k)\left|\theta(\vec k)\right|^2.
\label{equ30}
\end{equation}
Here $J(\vec k)$ can be simplified with the long-wave approximation,
\begin{equation}
J(\vec k) = \frac{1}{2}\sum_{\vec a} \left| 1 - e^{i\vec k \cdot
\vec a }\right|^2 \approx a^2k^2. \label{equ40}
\end{equation}

The relaxation dynamics of model A for the XY model can be described by the
Langevin equation,
\begin{eqnarray}
\frac{d \theta(t, \vec k)}{dt} & = & - \frac{1}{T} \frac{\partial H}{\partial \theta(t, \vec k)} + \epsilon(t, \vec k)
=-\frac{a^2k^2}{T}\theta(t, \vec k) + \epsilon(t, \vec k), \label{equ50}\\
\theta(t, \vec k)& = & \frac{a}{2\pi}\int dR e^{-i\vec k\cdot \vec R}\theta(t, \vec R). \label{equ55}
\end{eqnarray}
Here $\epsilon(t, \vec k)$ is a Gaussian white
noise with $\langle \epsilon(t, \vec k) \epsilon(t', \vec k')\rangle = 2 \delta(\vec k + \vec k') \delta(t
- t')$, and $\langle \cdots\rangle$ represents the statistical average.
Since the equation is linear, it can be exactly solved,
\begin{equation}
\theta(t, \vec k) = \int_{0}^{t}e^{-a^{2}k^{2}\left(t-t'\right)/T}\epsilon(t', \vec k) dt' + \theta(0, \vec k) e^{-a^{2}k^{2}t / T }.
\label{equ60}
\end{equation}
One may easily calculate
$\langle \theta(t, \vec k)\rangle = \theta(0, \vec k) e^{-a^{2}k^{2}t/ T}$
and
\begin{equation}
\langle |\theta(t, \vec k)|^{2}\rangle =
\frac{T}{a^2k^2}\left(1-e^{-2a^{2}k^{2}t/T}\right) +
|\theta(0, \vec k)|^{2}e^{-2a^{2}k^{2}t/T}.
\label{equ70}
\end{equation}

In Eq.~(\ref{equ60}), $\theta(t, \vec k)$
consists of two parts, the bulk one $F(t, \vec k)$ and the initial-condition dependent one
$G(t, \vec k)$,
\begin{eqnarray}
F(t, \vec k) =
\int_{0}^{t}e^{-a^{2}k^{2}\left(t-t'\right)/T}\epsilon(t', \vec k)dt', \quad
G(t, \vec k)  =  \theta(0, \vec k)e^{-a^{2}k^{2}t/T}.
\label{equ80}
\end{eqnarray}
From the inverse fourier transformation, one calculates $F(t, \vec R)$ and $G(t, \vec R)$, and then obtains
the complex magnetization,
\begin{equation}
M_x(t, \vec R) +i M_y(t, \vec R) = \left \langle e^{i\theta(t, \vec R)} \right \rangle  = \left
\langle e^{iF(t, \vec R)} \right \rangle e^{iG(t, \vec R)}.
\label{equ90}
\end{equation}
Using the cumulant expansion and the result
$\langle F(t, \vec R)\rangle= 0$, one deduces
\begin{eqnarray}
M_x(t, \vec R) &=& e^{- \langle F^2(t, \vec R) \rangle/2 } \cos(G(t, \vec R)), \nonumber \\
M_y(t, \vec R) &=& e^{- \langle F^2(t, \vec R) \rangle/2 } \sin(G(t, \vec R)).
\label{equ100}
\end{eqnarray}
One may calculate
\begin{eqnarray}
\langle F^2(t, \vec R)\rangle & = & \left(\frac{a}{2\pi}\right)^{2}\int d\vec k\int d\vec k'
\left\langle F(t, \vec k)F(t, \vec k')\right\rangle e^{i\left(\vec k+\vec k'\right)\cdot \vec R} \nonumber \\
& = & \left(\frac{a}{2\pi}\right)^{2}\int d\vec k\frac{T}{a^{2}k^{2}}\left(1-e^{-2a^{2}k^{2}t/T}\right) \nonumber \\
& = & \frac{T}{4\pi}\ln t + C,
\label{equ120}
\end{eqnarray}
where $C$ is an integral constant. Obviously the bulk part of $\theta(t, \vec R)$ is independent of $\vec R$.

Now it is important to calculate the initial-condition dependent part,
\begin{eqnarray}
G(t, \vec R)& = & \frac{a}{2\pi}\int d\vec k e^{i\vec k\cdot \vec R}G(t, \vec k).
\label{equ130}
\end{eqnarray}
For the ordered initial state, i.e., $\theta(0, \vec R) = \phi$, one calculates
$G(t, \vec R) = \phi$, and
\begin{equation}
M_x(t, \vec R) \propto t^{- T/8\pi}\cos \phi,
M_y(t, \vec R) \propto t^{- T/8\pi}\sin \phi.
\label{equ140}
\end{equation}
Under the long-wave approximation, $\eta = T/2\pi$ and $z = 2$ \cite{bra00a}.
Thus the power-law behavior of the magnetization, i.e., $M \propto t^{- \eta / 2 z}$, agrees
with that from the scaling theory and
renormalization-group calculations \cite{jan89,zhe98}.

For the spin-wave initial state, i.e.,
$\theta(0, \vec R)=2 \phi \cos(k_0 x)$ with $\vec R=(x, y)$,
\begin{eqnarray}
G(t, x)& = & \frac{a}{2\pi}\int d\vec k e^{i\vec k\cdot \vec R}G(t, \vec k)=2 \phi \cos (k_0 x) e^{-a^2k_0^2 t / T}.
 \label{equ220}
\end{eqnarray}
Since we only consider the case of small $k_0$, e.g., $k_0 \sim 1/L$,
the time dependence in Eq.~(\ref{equ220}) can be ignored.
Hence $G(t, x)$ is approximately equal
to $\theta(0, \vec R)$. Then
\begin{equation}
M_x(t, \vec R)  \propto  t^{- \eta/2z}\cos(2\phi \cos(k_0 x)),
M_y(t, \vec R)  \propto  t^{- \eta/2z}\sin(2\phi \cos(k_0 x)).
\label{equ230}
\end{equation}
Obviously the standard magnetization is zero.
But the local pseudo-magnetization is non-zero. In fact the local pseudo-magnetization is independent of $x$,
\begin{equation}
M_{p}(t, \vec R)  =  \vec{M}(t, \vec R)\cdot \vec{M}(0, \vec R) \propto  t^{- \eta/2z}.
\label{equ240}
\end{equation}
This power-law behavior is in agreement with the scaling form in Eq.~(\ref{equ18}).

How is the dynamic relaxation of a vortex state? In this case,
$\theta(0, \vec R)=\varphi(\vec R)+\pi/2$, and $\varphi(\vec R)$
is the polar angle of $\vec R$. According to
Eqs.~(\ref{equ80}) and (\ref{equ130}), one can calculate,
\begin{equation}
G(t, \vec r) = \frac{T}{4\pi t} \int d\vec R \theta(0, \vec R) e^{-(\vec R-\vec r)^2 /(4a^2t/T)}\ .
\label{equ150}
\end{equation}
Let us define $\vec R'=(\vec R-\vec r)/\sqrt{4a^2t/T},
\vec {r'} = \vec r/\sqrt{4a^2t/T}$, then
\begin{equation}
G(t, \vec {r'})  =   \frac{1}{\pi} \int dR' R' e^{-R'^2} \int d\varphi(R') \theta(\vec {R'}+ \vec {r'},0).
\label{equ170}
\end{equation}
Due to the rotational symmetry, $G(t, \vec {r'}) - \theta(0, \vec {r'})$ is
independent of $\varphi (\vec {r'})$. To simplify the calculations, we set $\vec {r'} = (r', 0)$, i.e., $\varphi(\vec {r'})=0$, thus
\begin{eqnarray}
G(t, \vec{r'}) - \theta(0, \vec {r'}) & = &  \frac{1}{\pi} \int dR'
R' e^{-R'^2}
\int d\varphi(R') \varphi(\vec {R'}+ \vec {r'}) \nonumber \\
& = & \frac{1}{\pi} \int_{0}^{\infty} dR' R' e^{-R'^2}
\int_{-\pi}^{\pi} d\varphi(R') \Theta(\varphi(R'), r'/R'),
\label{equ180}
\end{eqnarray}
and $\Theta(\varphi(R'), r'/R')$ is introduced for convenience,
\begin{equation}
\Theta(\varphi(R'), r'/R')= \arctan
[\cos\varphi(R')+ \frac{r'}{R'}, \sin\varphi(R')].
\label{equ185}
\end{equation}
Here $\arctan(x,y)$ is defined in the interval $[-\pi, \pi]$ by $\tan [\arctan(x,y)]=y/x$.

In fact, $\varphi(R')$ is the polar angle of $\vec R'$, and $\Theta(\varphi(R'), r'/R')$
is that of $\vec R'+\vec {r'}$. $\vec {r'}$ is the fixed site we are looking at, and $\vec R'$ is the site we should integrate.
For $R' < r'$, $|\Theta(\varphi(R'), r'/R')|<\pi/2$, and $\Theta(\varphi(R'), r'/R')$ is a single-valued
odd function of $\varphi(R')$. The integration over $\varphi(R')$ in Eq.~(\ref{equ180}) is zero. For
$R' \geq r'$, $\Theta(\varphi(R'), r'/R')$ is
still odd, but not single-valued at $\varphi(R')= \pm \pi$.
When $\varphi(R')\rightarrow \pm \pi$, $\Theta(\varphi(R'), r'/R') \rightarrow \pm \pi$. However,
$\varphi(R')=\pm \pi$ correspond to a same spatial point.
Since $\Theta(\varphi(R'), r'/R')$ is multiple-valued at $\varphi(R')= \pm \pi$,
the integration over $\varphi(R')$ in Eq.~(\ref{equ180}) can be non-zero.

From the above analysis, the integration
in Eq.~(\ref{equ180}) can be rewritten as
\begin{equation}
G(t, \vec{r'}) - \theta(0, \vec {r'}) = \frac{1}{\pi}
\int_{r'}^{\infty} dR' R' e^{-R'^2} \int_{-\pi}^{\pi} d\varphi(R')
\Theta(\varphi(R'), r'/R').  \label{equ250}
\end{equation}
Let us focus on the integration over $\varphi(R')$. Since $\Theta(\varphi(R'), r'/R')$ is an odd function,
non-zero contribution only comes from a small interval around $\varphi(R')=\pm \pi$, and
\begin{eqnarray}
 & \quad & \int_{-\pi}^{\pi} d\varphi [\Theta(\varphi + \epsilon) -
 \Theta(\varphi - \epsilon)]= 2 \int_{-\pi}^{\pi} d\varphi \Theta (\varphi),
\label{equ260}
\end{eqnarray}
where $\epsilon$ is a small angle to be determined. Substituting Eq.~(\ref{equ260}) into Eq.~(\ref{equ250}),
\begin{eqnarray}
G(t, \vec{r'}) - \theta(0, \vec {r'}) & = & \frac{1}{2\pi} \int_{r'}^{\infty} dR' R' e^{-R'^2}\int_{-\pi}^{\pi} d\varphi [\Theta(\varphi + \epsilon) - \Theta(\varphi - \epsilon)]  \nonumber \\
& = & \frac{1}{2\pi} \int_{r'}^{\infty} dR' R' e^{-R'^2}\int_{-\pi}^{\pi} d\Theta \frac{\Delta}{R'} \nonumber \\
& = & \pm \Delta \int_{r'}^{\infty} dR' e^{-R'^2}.
\label{equ270}
\end{eqnarray}
Here $2\epsilon=\Delta/R'$, and $\oint d\Theta=\pm 2\pi$ dependent on the existence of a vortex or anti-vortex.

The pseudo-magnetization can be calculated by $M_{p}(t, \vec {r'}) =
t^{-\eta / 2z}\cos(G(t, \vec {r'}) - \theta(0, \vec {r'}))$. When $r'
\rightarrow \infty$, it approaches the bulk behavior $M_{p}(t, \vec {r'}) = t^{-\eta/2z}$. When $r' \rightarrow 0$,
$M_{p}(t, \vec {r'}) = t^{-\eta/2z}\cos(\Delta \sqrt{\pi} / 2)$. Due to the symmetry, or from our numerical simulations,
$M_{p}(t, 0)$ should be zero. Hence the parameter $\Delta = \sqrt{\pi}$ is determined.
The finial form of the pseudo-magnetization is then derived
\begin{equation}
M_{p}(t,  {r'}) = t^{-\eta / 2z}\sin(\sqrt{\pi}
\int_{0}^{r'}dR'e^{-R'^2}), \label{equ210}
\end{equation}
where $r' = r / \sqrt{4a^2t/T} \propto r/t^{1/z}=s$.

The linearized long-wave approximation in Eq.~(\ref{equ420}) is good for the dynamic relaxation of the spin-wave state, 
for the dynamic effect of the vortex-pair creation and annihilation is weak. 
For the dynamic relaxation of the vortex state, however, the linearized long-wave approximation
could not catch the logarithmic correction in the core of the vortex. 
Additionally, a phenomenological technique has been used in the derivation of Eqs.~(\ref{equ270}) and (\ref{equ210})
to obtain the scaling form in the core of the vortex.

\begin{table}[h]\centering
\begin{tabular}[t]{ c | c | c | c | c | c }
 \hline
  & STD  \cite{zhe03a}  &  DW  \cite{he09}&      Spin-wave &    Vortex  &    Disorder   \\
\hline
$\eta$        &   0.234(2)      & 0.234(2)       &   0.234(2)          & 0.234(2)    &            \\
$\eta_0/2$    &                       &   1.00(1)   &                 & 0.99(2)     & 0.80(2)    \\
$d_0$         &                     &   1.99(1)   &               & 0.97(4)     & 0.78(3)       \\
$z$           &   2.01(1)            &   1.99(4)  &   1.97(4)            & 1.98(4)     &          \\
 \hline
\end{tabular}
\caption{Critical exponents estimated from the relaxation dynamics of the vortex state and spin-wave state,
in comparison with those from the domain-wall (DW) dynamics and short-time dynamics (STD).
In the last column, quenched disorder is located at the vortex center. } \label{t1}
\end{table}

\end{document}